\documentstyle[multicol,aps,prl,epsf,graphics,psfig]{revtex}

\begin{document}

\draft
\preprint{}

\title{Superfluid-spiral state of quantum ferrimagnets in magnetic field} 

\author{M.~Abolfath$^{\text a}$, and A.~Langari$^{\text b,c}$}

\address{ 
$^{\text a}$  Department of Physics and Astronomy,
University of Oklahoma, Norman, OK 73019-0225} 
\address{ 
$^{\text b}$
Max-Planck-Institut f\"ur Physik komplexer Systeme, N\"othnitzer Strasse 38,
D-01187 Dresden, Germany }
\address{ 
$^{\text c}$ Institute for Advanced Studies in Basic Sciences, 
Zanjan 45195-159, Iran} 
\date{\today}
 
\maketitle 
 
\begin{abstract} 
\leftskip 2cm 
\rightskip 2cm 
We study the phase diagram of one-dimensional quantum ferrimagnets  
by using a numerical exact diagonalization of a finite size system
along with a field-theoretical non-linear $\sigma$ model  
of the quantum ferrimagnets  
at zero temperature and its effective description in the presence of 
the external magnetic field in terms of the quantum XY-model.
The low- and the high-field phases correspond respectively  
to the classical N\'eel and the fully polarized ferromagnetic states  
where in the intermediate magnetic field ($h_{c1} < h < h_{c2}$), 
it is an XXZ+h model with easy plane anisotropy, which possess the
spiral (superfluid) states that carry the dissipationless
spin-supercurrent. 
We derive the critical exponents, and then will study the stability of the 
XY spiral state 
against these spin-supercurrents and the hard axis fluctuations.
We will show a first order phase transition from the easy plane spiral 
state to a saturated ferromagnetic state occurs at $h=h_{c2}$ if the 
spin-supercurrent reaches to its critical value. 
\end{abstract} 
 


\begin{multicols}{2}
\section{Introduction}
Recently, antiferromagnetically coupled mixed-spin chains with an 
alternating 
array of two kinds of spins have attracted interest among 
researchers \cite{devega,pati,Yammamoto,Kolezhuk,Homayoun}. 
Integrable models of mixed-spin antiferromagnetic chains were constructed 
by de~Vega and Woynarovich \cite{devega} and the 
simplest case of such chains with spins $S=1$ and $1/2$ were subsequently 
studied \cite{pati}. Since these integrable models are exactly solvable, 
they are very useful for studying (quantum) statistical mechanical 
properties. 
Although ferrimagnetic spin  chains exhibit 
both ferromagnetic and antiferromagnetic features, they show some 
peculiar, 
and sometimes surprising,  features uncharacteristic of either the 
ferromagnet or the antiferromagnet---an example being the 
existence of gapless excitations with very small correlation length. 
It is important to understand these features more clearly.  
In this paper we first present a numerical study on the quantum 
ferrimagnetic spin chain in magnetic field by 
an exact diagonalization of a finite size system. 
We show that the low- and the high-field phases correspond respectively  
to the classical N\'eel and the fully polarized ferromagnetic states  
where in the intermediate magnetic field ($h_{c1} < h < h_{c2}$), 
it is an XXZ+h model with easy plane anisotropy \cite{Kolezhuk,Homayoun},
and the critical exponents are derived numerically.
Then the physical properties of the quantum ferrimagnets 
in the presence of the uniform magnetic field, as described  
by a phenomenological field theory
based on a continuum non-linear $\sigma$ model (NL$\sigma$M)
formulation are examined further
and will explore its novel features. 
When $h \geq h_{c1}$, the quantum ferrimagnets can be described
by an easy-plane anisotropy state where the broken $U(1)$ symmetry
of spins corresponds to a topological spiral (superfluid) state.
A metastable state with a non-zero topological spin-supercurrent $J_s$,
is the superfluid state of the order parameter in XY-plane.
Such possibility, may be led to the remarkable spin-dependent 
transport phenomenon in the ferrimagnetic chains, i.e.,
the spin-supercurrent is carried collectively (rather than by quasiparticles).
Because the spin current is non-zero when the system is in equilibrium,
it flows without dissipation.
The spiral state can be characterized by a wave-vector $Q$, i.e., the soliton
(saw tooth) lattice 
spacing $a=2\pi/Q$ is a length scale through which the in-plane phase
$\varphi$ of the order parameter changes by $2\pi$.
Hence, the spin-supercurrent is characterized by these wave-vectors, e.g.,
$J_s \propto Q$.
In analogy with the dissipation mechanism of the supercurrent
in a one-dimensional superconductors \cite{Langer}, at large
$J_s$, hard-axis fluctuations of the order parameter becomes
unstable and a first order phase
transition to a uniform in-plane state ($\varphi={\rm cte}$)
takes place, and the long range order of the soliton lattice destroys.
By overcoming the nucleation energy barrier of the order parameter vortices,
a first order phase transition from easy plane spiral state
to a saturated ferromagnetic state occurs at $h=h_{c2}$, depends on 
$J_s$, and $Q$. 
This transition turns out to a continuous cross-over if the
initial state of the system at $h=h_{c1}$ holds the zero spin-supercurrent
(a uniform easy plane with $Q=0$). 
The possibility of the existance of the superfluid
phase and crossing onto a saturated ferromagnetic state 
(the dependence of $h_{c2}$ on $Q$, and $J_s$)
are discussed in some detail.

The zero-temperature quantum ferrimagnetic chain is defined by
$H=J\sum_{<ij>} S_i \cdot s_j - h \sum_i(S_i+s_i)$ where $S \neq s$.
The low energy physics of the quantum ferrimagnets 
in the presence of the external magnetic field can be obtained
by a 1-dimensional NL$\sigma$M.
The external magnetic field couples with 
${\bf n}$, the N\'eel order parameter
\begin{equation}
{\cal L}_h= i M_0 {\bf A}({\bf n}) \cdot \partial_\tau {\bf n} 
+ {\cal L}_{\rm NL\sigma M+h} + {\cal L}_{\rm top} - M_0 \vec{h}\cdot{\bf n}, 
\label{lag_h}
\end{equation}
where 
\begin{equation}
{\cal L}_{\rm NL\sigma M+h} = \frac{1}{2g}\left(v_s(\partial_x {\bf n})^2
+ \frac{1}{v_s}[2i\vec{h} + {\bf n}\times\partial_\tau{\bf n}]^2\right).
\label{NLSM_h}
\end{equation}
The first term in Eq.~(\ref{lag_h}) is the usual (dynamical) 
Berry's phase of a quantum ferromagnet, with $M_0 \equiv |S-s|/a_0$ the 
magnetization per unit cell (pair of sites). 
Here ${\cal L}_{\rm top}$ is the topological term,
$g=4/(s+S)$, and $v_s=4a_0J s S/(s+S)$ \cite{Homayoun}.
It is this term that results in the ferromagnetic branch of 
the spin waves and corresponds to the trajectory of spin over a closed 
orbit on the unit sphere in the presence of a unit magnetic monopole at 
the center. 
The contribution of the first term in ${\cal L}_h$ is equivalent to 
the area enclosed by this trajectory and since either of the smaller or 
the larger enclosed areas on the unit sphere must lead to the same Berry's 
phase, 
the magnetic moment per unit cell, 
i.e. $2 M_0 a_0$, must be quantized with integral values \cite{read}. 
Setting $M_0 = 0$ in action (\ref{lag_h}) follows to the usual $O(3)$
NL$\sigma$M in the magnetic field \cite{Nicopoulos}, applicable to
the Heisenberg antiferromagnets. 
Similar to zero magnetic field \cite{Homayoun}
we can find the spin-wave modes.
At $h=0$ the ferrimagnetic spin-waves consist of both gapless (ferromagnetic)
and gapped (antiferromagnetic) modes, and at $T=0$ 
the low (high) energy physics of quantum ferrimagnets is effectively  
like that of a ferromagnet (antiferromagnet) 
which is formed by the chain 
of (dimerized) unit cells with magnetic moment $M_0(=|S - s|)$. 
Applying an external magnetic field, $h$,  
develops a gapped ferromagnetic spin-waves; in which case  
the energy cost for the ferromagnetic transitions is proportional to the 
Zeeman splitting factor. 
Unlike to ferromagnetic mode, the effect of the external magnetic field 
is to suppress the antiferromagnetic gap. 
Clearly, this reveals similarities between the ferrimagnets 
and the integer spin Heisenberg antiferromagnets in magnetic field
\cite{Sakai}.
The ground state of  
the ferrimagnet corresponds to the {\em staggered}  configuration of spins,   
unless $h \geq h_{c1}=2J|S-s|$. 
At this point the staggered 
state becomes unstable against the non-collinear {\em spin-flop} phase  
(the partially polarized state) of  
the spins when the spectrum becomes soft at $k=0$. When the external magnetic 
field exceeds $h_{c2}$, the system will be in a saturated 
ferromagnetic phase, with a quantized magnetization per unit cell. 
The $h_{c2} \equiv 2J(S+s)$ is  
obtained by using the dispersion relation of the spin waves based on the fully  
polarized state of the ferrimagnets.  
It is the lower-bound of the external magnetic 
field, in the sense that the spin waves of the fully polarized state
become soft at $k=0$.

The outline of this paper is as follows : In next section we will 
present the numerical computations of Lanczos method to compare the behavior 
of the correlation functions on the plateau and between two plateaux. 
We also derive the critical exponents, and
the effective Hamiltonian for the latter region by
using a Quantum Renormalization Group (QRG) approach. 
The main part of this paper, the spiral (superfluid) state of the quantum 
ferrimagnets is presented in section III.
The details of the collective modes in the intermediate
magnetic phase is given as the appendix in Sec. IV.
 
\section{Numerical results}
In order to have more accurate physical picture, we present the
numerical results, by applying a Lanczos method to each $S_z$-sector of
Hilbert space from  $S_z=\frac{N}{2}(S - s)$   
to $S_z=\frac{N}{2}(S + s)$. 
Here $N(=20)$ is the number of sites that is used in the exact diagonalization.

\begin{figure} 
\epsfxsize=7.5cm \epsfysize=9.0cm {\epsffile{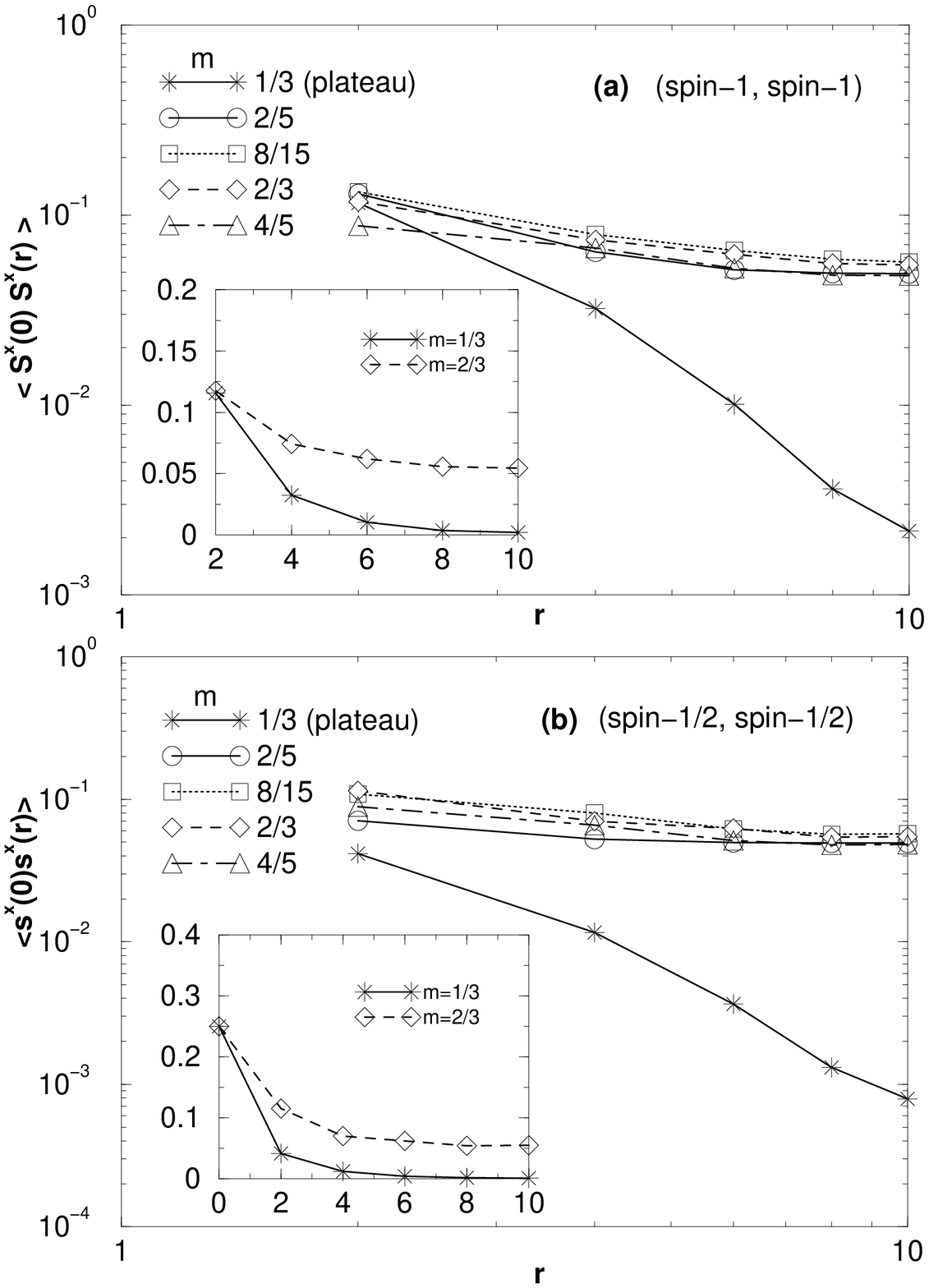}}
\epsfxsize=7.5cm \epsfysize=4.5cm {\epsffile{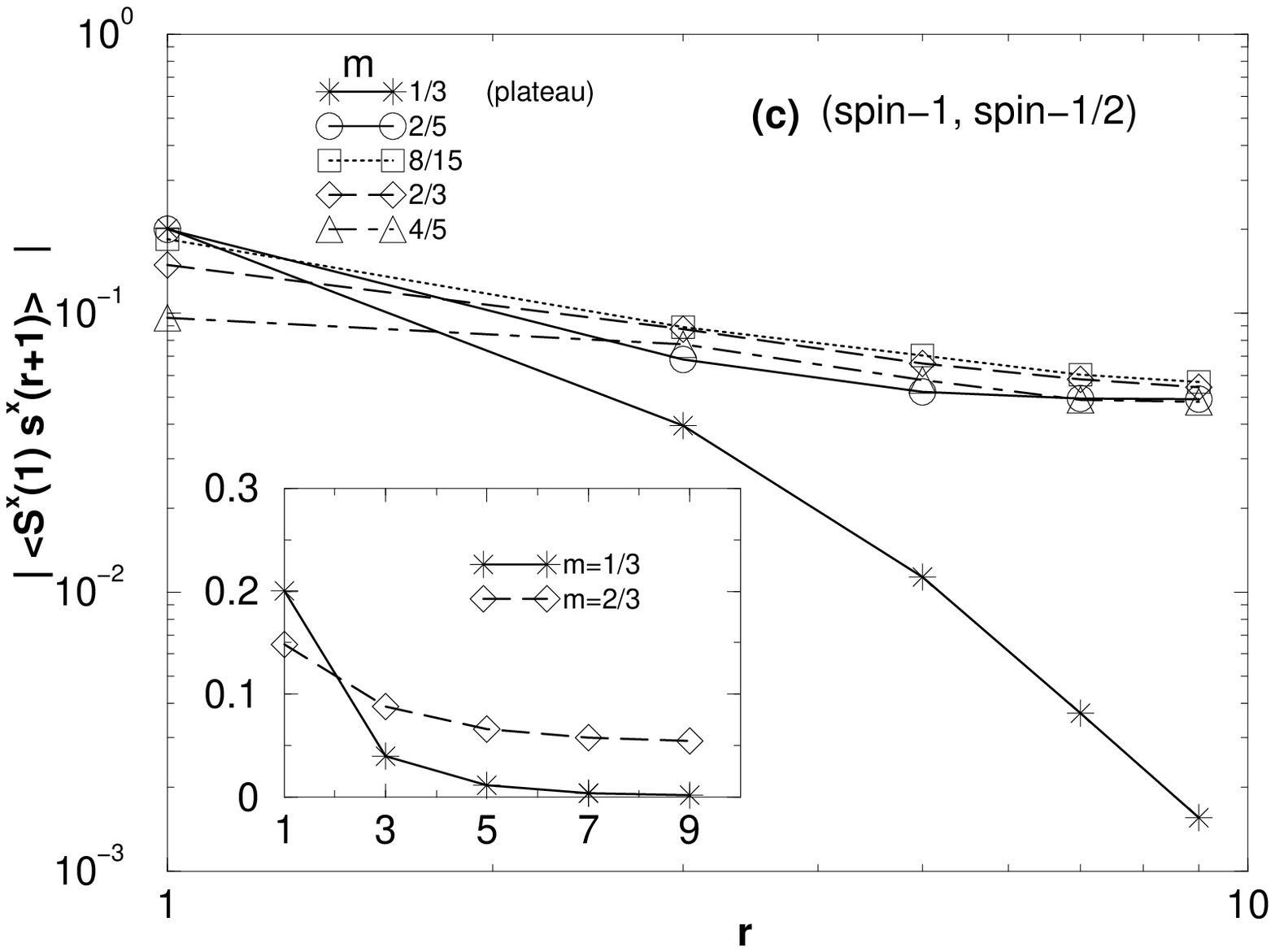}}
\caption{The log-log plot of in-plane spin-spin correlation 
function is shown within the
plateau ($m=1/3$) and within the partially spin polarized state 
($m=2/5, 8/15, 2/3, 4/5$).
The former shows the exponential decay while in the 
latter region the power law behavior of correlation function is
seen. The correlation on sublattice A (spin-1, spin-1) is shown in (a), 
on sublattice B (spin-1/2, spin1/2) in (b) and between the two sublattices
A (spin-1) and B(spin-1/2) in (c). In (c) spin-1 is fixed and the position
of spin-1/2 is running to show the correlations. 
In each part an inset represents the correlation for 
$m=1/3$ and $2/3$ in a normal scale. The former approaches to zero
exponentially, and the latter falls off algebraically.
} 
\label{fig2} 
\end{figure}

A curve for magnetization vs. magnetic field for a chain of
$(1/2,1)$-ferrimagnet has been given in Ref. \cite{ferriladder}.
An extrapolation to $N \rightarrow \infty$ on exact diagonalization 
calculation shows there are two plateaux of magnetization
below $h_{c1}=\Delta_0=1.7589 J$ and above $h_{c2}=3J$, 
associated with the magnetization $m\equiv\frac{M}{S+s}=1/3$ and $m=1$. 
Here $\Delta_0$ is the energy gap between the Ferro- and
Antiferromagnetic spin wave modes of this model. 
In Fig. \ref{fig2} the numerical results of 
the in-plane spin-spin correlations is presented for  
a point on the plateau (at $m=1/3$) and some intermediate points
between two plateaux, i.e., $m=2/5, 8/15, 2/3$ and $4/5$. 
This exhibits the in-plane spin-spin correlation functions
within the intermediate magnetic field region which falls off as power law,
and manifests the critical behavior.

\noindent
Thus the in-plane  correlation is expected to have the asymptotic
form $\langle S^x(0) S^x(r) \rangle \sim r^{-\eta}$.
For instance we have calculated this exponent for m=$2/3$ by using 
data of exact diagonalization of a chain with length $N=20$.
Since there are correlations between different types of sublattices we find 
an exponent for each case. For correlations on sublattice A 
(Fig.\ref{fig2}-(a)) and m=$2/3$ we obtained $\eta=0.44 \pm 0.01$ and 
for sublattice B (Fig.\ref{fig2}-(b)) we have $\eta=0.42 \pm 0.04$.
A similar behavior is seen for the correlations of different sublattices 
A and B (Fig.\ref{fig2}-(c)) where $\eta=0.47 \pm 0.01$.
To have more qualitative picture of the transient region
between two plateaux, we can use a quantum renormalization group. 
To perform this, we choose two adjacent spins $(S=1,s=1/2)$ as the building
block.
The block Hamiltonian is then $H_B = J {\bf S}\cdot{\bf s} - h(S_z+s_z)$.
For low field limit (e.g. $h < h_{c1}$) the lowest lying states of $H_B$ are
a spin-1/2 doublet. This comes out with an effective Hamiltonian of
ferromagnetic Heisenberg chain with $S=1/2$ in magnetic field, corresponds 
to the $m=1/3$ plateau. 
For high field limit $h > 3J/2$ where two states 
$|S_T =1/2, S^z_T=1/2\rangle$, and $|S_T =3/2, S^z_T=3/2\rangle$
are nearly degenerate, we arrive with a spin-$1/2$ antiferromagnetic
$XXZ+h$ Hamiltonian. 
The effective Hamiltonian is
\begin{equation}
H_{\rm eff}=\frac{2J}{3}\sum_{n=1}^{N/2} 
(\tau^\perp_n \cdot \tau^\perp_{n+1} + \Delta \tau^z_n \tau^z_{n+1}) 
- h' \sum_{n=1}^{N/2} \tau^z_n
\label{heff}
\end{equation}

\noindent where $\tau$ is spin-1/2 operator, 
$\Delta=1/3, h' = \frac{3}{2}h - \frac{19}{6}J$.
It is known such model Hamiltonian
in the presence of a magnetic field ($h$) has a critical line which separates
the partially polarized phase from the fully polarized one
\cite{yang}. The magnetization of the 
ferrimagnetic chain ($m$)
is related to the magnetization of $XXZ+h$ model by $m=1+m_{XXZ+h}$ which leads
to the same behavior as 
the numerical results
but with some discrepancies for
the critical fields ($h_{c1}=1.5J\;, \;h_{c2}=3J$).
However one may continue the quantum RG procedure for the $XXZ+h$ model and 
obtain that the RG flow for the partially polarized phase (between the 
two plateaux) goes to the isotropic $XY$ fixed point \cite{langari}. 


\section{Superfluid phase}
\begin{figure}
\center
\epsfxsize 6.0cm \rotatebox{-90}{\epsffile{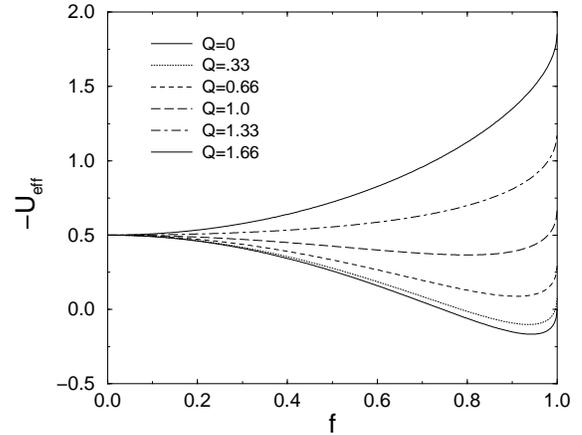}}
\vskip.5cm
\caption{
$-U_{\rm eff}(f)$, the effective potential of the spiral state
vs. $f$ is plotted for $h^*=J$. 
The minimum energy solution at $Q=Q^*$ is $f=0$ where the 
spiral state destroys. 
}
\label{Fig_Ueff}
\end{figure}

By generalizing the NL$\sigma$M$+$h
we propose a phenomenological field theory which can present 
the long wave length limit of the quantum ferrimagnets within
$h \geq h_{c1}$
\begin{equation} 
E = \int dx \left[\frac{1}{2} \rho_s^\perp (\partial_x n_\perp)^2
+ \frac{1}{2} \rho_s^z (\partial_x n_z)^2 
+ \beta n_z^2 - h^* n_z\right],
\label{eq9}
\end{equation}
where we applied the spin coherent state formalism in Eq.(\ref{eq9}).
Here 
$h^* = h - h_{c1}$ is the effective magnetic field.  
$\rho_s^\perp$ and $\rho_s^z$ are in-plane and out of plane
spin stiffness. $\beta > 0$ gives an easy plane anisotropy.
To study a superfluid phase, it is convenient to introduce
the following variational solutions
${\bf n}(x) = (\sin\theta\cos\varphi, \sin\theta\sin\varphi,
\cos\theta$), where $\theta=\theta(x)$ and $\varphi = Qx$. 
Here $Q$ is a spiral state wave vector responsible for 
the superfluid phase slip,
and $a=2\pi/Q$ is the soliton lattice spacing, i.e., the phase 
$\varphi$ changes by $2\pi$ along the chain.
Clearly $\tilde{n}_z=\cos\theta$ is the uniform solution
($\theta$ is a constant) and 
$\tilde{E} = \rho_s^\perp Q^2 (1-\tilde{n}_z^2)/2
+ \beta \tilde{n}_z^2 - h^* \tilde{n}_z$
is the corresponding energy per unit length.
Note $n_z$ is the momentum density conjugate to field variable $\varphi$,
and the Hamiltonian (\ref{eq9}) gives a linearly dispersing collective mode,
associated with the $U(1)$ symmetry breaking phase, i.e., the superfluid phase.
It is easy to check how the classical solutions
and the fluctuating modes can be derived by Eq.(\ref{eq9}).
For example $d\tilde{E}/d\tilde{n}_z = 0$ leads to 
$\tilde{n}_z=h^* /\tilde{K}_{zz}$ where 
$\tilde{K}_{zz} \equiv d^2\tilde{E}/d\tilde{n}_z^2 = 
2\beta - \rho_s^\perp Q^2$ is the energy gap of 
the out-of-plane modes at zero wave vector.
Then  
$J_s = (1/\hbar) d\tilde{E}/dQ = \tilde{n}^2_\perp\rho_s^\perp Q/\hbar$
is the gauge invariance
topological spin-supercurrent density carried along the $x$-direction.
Furthermore, it is necessary to study the stability 
of the superfluid phase against the quantum fluctuations. 
This governs with the in-plane fluctuations
$K_{\varphi\varphi}(k)=2 \rho_s^\perp \tilde{n}^2_\perp k^2$, and
the out of plane fluctuations
\begin{eqnarray} 
K_{zz}(k) &=& \rho_s^\perp \tilde{n}^2_z (k^2 + Q^2) - 
\rho_s^\perp \tilde{n}^2_\perp Q^2
\nonumber \\ &&
+ \rho_s^z \tilde{n}^2_\perp k^2 
+ 2\beta(\tilde{n}^2_\perp-\tilde{n}^2_z)
+ h^* \tilde{n}_z.
\label{eq7}
\end{eqnarray} 
As one can see $\tilde{K}_{zz}=K_{zz}(k=0)$ at $h^*=0$.
The gapless linear superfluid mode can be obtained by
$\hbar\omega = 2 \sqrt{K_{\varphi\varphi} K_{zz}}$.
The details of the collective modes calculation is presented in Appendix.
Through out this formulation, the zero temperature phase diagram 
of this system can be obtained.
At $h^*=0$ where $\tilde{n}_z =0$ the out of plane instability
occurs at $Q_c = \sqrt{2\beta/\rho_s^\perp}$.
At this point
the energy gap of the $zz$-modes vanishes
and the $O(3)$ symmetry of the Hamiltonian is restored
(the dispersion relation of the collective modes become imaginary at 
small $k$).
It follows the transition to a uniform solution must occurs at this point 
since $\pi_1(S^2) = 0$ where the linear solution can be considered
as a map from the compactified
physical space to the equator of the order parameter
space $S^2$. 
Such instability in order parameter space is induced by the local fluctuations 
of the order 
parameter, out of the equator toward the north pole of the order
parameter space $S^2$.
When $J_s$ is close to its critical values ($J_{sc}$),
the local fluctuations are very strong, and it is likely
the order parameter pass the north pole, and 
the phase $\varphi$ becomes singular, i.e., a vortex
in $\varphi$ is nucleated, and the phase winding is lost.
The effect of $h^*$ is pushing the spins to
be aligned along z-axis, e.g., $\tilde{n}_z=1$ at large enough $h^*$.
We notice if $h > h_{c1}$, system
undergoes onto the saturated ferromagnetic phase by increasing
the spin-supercurrent density. 
This happens at $Q^* = \sqrt{(2\beta-h^*)/\rho_s^\perp} < Q_c$,
corresponding to $J_s=0$,
before the outset of the easy-plane uniform state. 
To study the nucleating mechanism of the vortices, we implement 
a mechanical analogy to the classical field theory.
This formalism has been developed by Langer, and Ambegaokar \cite{Langer}
for one dimensional superconductors.
To avoid the complexity let us
neglect $\rho_s^z$ in energy functional (\ref{eq9}) from now on.
To start this calculation, we make the following 
transformations $n_x(x) = f(x) \cos\varphi(x)$, 
$n_y(x) = f(x) \sin\varphi(x)$, and $n_z(x) = \sqrt{1-f^2(x)}$
(where $f \equiv n_\perp$), and
\begin{eqnarray} 
E[f] &=& \int dx \left(\frac{\rho_s^\perp}{2}\left[(\partial_x f)^2
+ \frac{L^2}{f^2}\right] \right. \nonumber \\ && \left.
+ \beta(1-f^2) - h^* \sqrt{1-f^2}
\right),
\label{eq10}
\end{eqnarray}
where $L$ is the momentum conjugate to $\varphi$ and it is proportional
to the spin-supercurrent density $J_s$, because $\partial_x \varphi = L/f^2$
is the solution of the $\delta E[f,\varphi]/\delta\varphi=0$. 
It is straightforward to show how $\delta E[f]/\delta f=0$ yields
\begin{mathletters}
\label{eq14}
\begin{equation} 
x - x_0 = \int_{f(x_0)}^{f(x)} \frac{df}{2\sqrt{E_{\rm eff} - U_{\rm eff}(f)}},
\end{equation}
\begin{equation}
\varphi(x) -  \varphi(x_0) = L~\int_{f(x_0)}^{f(x)} \frac{dx}{f^2},
\end{equation}
\begin{equation}
- U_{\rm eff}(f) = \frac{\rho_s^\perp L^2}{2f^2} + \beta(1-f^2)
- h^* \sqrt{1-f^2}.
\end{equation}
\end{mathletters}
Uniform solutions ($f=\tilde{f}={\rm cte}$) are one of the solutions 
of Eqs. (\ref{eq14}). It leads to $\varphi = Qx$ and $L=Q \tilde{f}^2$.
The energy potential associated with the uniform solutions are
depicted in Fig. \ref{Fig_Ueff} where
$-U_{\rm eff}(f)$ vs. $f$ is plotted 
for $h^*=J$ for different $Q$'s, or equivalently for different
spin-supercurrent $J_s$. 
Here $\rho_s^\perp=2J/3$, and $\beta=J$.
The effective potential, $-U_{\rm eff}$, has one minimum at 
$\tilde{f}=1-[h^*/(2\beta - \rho_s^\perp Q^2)]^2 \neq 0$ if $Q \leq Q^*$.

\begin{figure}
\center
\epsfxsize 6.0cm \rotatebox{-90}{\epsffile{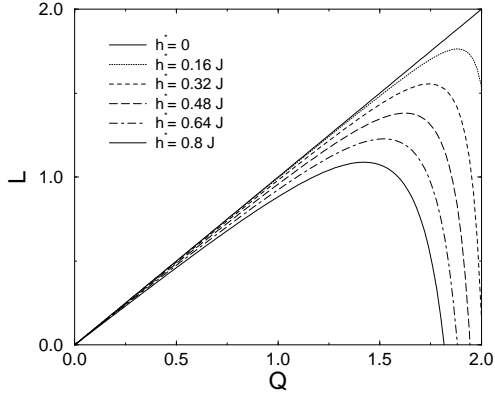}}
\vskip.5cm
\caption{$L$ vs. $Q$ is plotted for different $h^*$.
For any $h^*$ there is a maximum spin-supercurrent $J_s=J_s^*~(< J_{sc})$ 
which can pass through the system at wave 
vector $Q'$. $J_s^*$ is a monotonically decreasing function of
$h^*$. At $Q=Q^* \neq 0$, the spin-supercurrent density vanishes ($L=0$).
}
\label{L.Q}
\end{figure}

\noindent
This minimum energy solution is disappeared 
if $Q > Q^*(=\sqrt{(2\beta-h^*)\rho_s})$,
and is replaced by another minimum at $\tilde{f}=0$.
At $Q^*$ a first order phase transition 
from $\tilde{f} \neq 0$ onto $\tilde{f}=0$ 
takes place by nucleating of a vortex.
One can follow the transition between these
minimum energy solutions by changing the parameters, $Q$ and $h^*$,
as illustrated in Fig. \ref{Fig_Ueff}.
The saddle point solutions can be derived by setting 
$d^2U_{\rm eff}/d\tilde{f}^2=0$ at the extremum of $U_{\rm eff}(\tilde{f})$.
It follows at $Q^* = \sqrt{(2\beta-h^*)/\rho_s^\perp}~({\rm or~
equivalently} J_s=0)$, 
the local minima at finite $\tilde{f}$ is disappeared and
$\tilde{n}_z=1$ becomes a unique minima of $U_{\rm eff}$.
Moreover, the {\rm cross-over} from the uniform state [a broken $U(1)$ 
symmetry state with zero spin-supercurrent] to a saturated ferromagnetic 
state occurs at $h^*=2\beta=2J$
($Q^*=0$), consistent with the exact diagonalization results ($h_{c2}=3J$).
Obviously at $h=h_{c1}$ and $Q=Q_c~(=\sqrt{2\beta/\rho_s})$, the sign of 
curvature of $-U_{\rm eff}$ changes, and therefore
$Q_c$ can be considered as the saddle point of 
$-U_{\rm eff}$, at which the spiral state becomes unstable against the
hard-axis fluctuations
and a transition to a uniform solution occurs. 
This is seen if the spin-supercurrent $J_s$ arrives to its critical value,
$J_{sc}$.
As was mentioned earlier, the mechanism for such transition
is nucleating of a vortex in the order parameter space.
For a given $Q$ and $h^*$,
it is straightforward to show $\tilde{E}(Q) = \rho_s^\perp Q^2/2 - 
h^{*2}/(2\beta - \rho_s^\perp Q^2)$. 
Recalling $J_s=(1/\hbar)d\tilde{E}(Q)/dQ$
leads to $J_s=\rho_s^\perp\tilde{f}^2 Q/\hbar$
($J_s=\rho_s^\perp L/\hbar$) which is the spin-supercurrent density and 
$\tilde{n}_z=h^*/(2\beta - \rho_s^\perp Q^2)$
is the classical solution of Eq.(\ref{eq14}).
The dependence of $L(= Q[1-h^{*2}/(2\beta - \rho_s^\perp Q^2)^2])$ with 
respect to $Q$ for various $h^*$ is illustrated in Fig. \ref{L.Q}. 
It follows that $J_s$ vanishes at $Q^* = \sqrt{(2\beta-h^*)/\rho_s^\perp}$
where the system crosses to the saturated ferromagnetic phase.
The maximum spin-supercurrent $J_s^*~(< J_{sc})$, which can pass through 
the system at finite wave vector $Q$ (and $h^* \neq 0$),  
is a monotonically decreasing function of $h^*$.

In conclusion, we have predicted the intermediate magnetic field
region of ferrimagnets can support the dissipationless flow of the  
spin-supercurrents, which may be observed by the advance techniques
of {\em spintronics}. 
The idea of the collective spin-supercurrent transport which has been 
presented in this paper is completely general, and can be
applicable to any 1-dimensional spin system with easy
plane anisotropy. The intermediate magnetic field phase of the ferrimagnets  
is one example.
Other example of this kind is the {\em isospin}-supercurrent transport of 
the quantum Hall bars in the bilayer electron systems. \cite{Allan} 

\section{Appendix}
Here we present the detailed calculation of the collective modes in the
intermediate magnetic field phase. 
The effective Hamiltonian in terms of the in-plane and the out of plane
fluctuations is given by 
\begin{eqnarray} 
H_f[\varphi, n_z] &=& \frac{1}{2} \sum_q \varphi(-q) 
K_{\varphi\varphi}(q) \varphi(q) \nonumber \\ &&
+ \frac{1}{2}\sum_q n_z(-q) K_{zz}(q) n_z(q),
\end{eqnarray}
where $K_{\varphi\varphi}=\delta^2 E/\delta\varphi^2$ and
$K_{zz} = \delta^2 E/\delta n_z^2$.
The equation of motion for the field variables $\varphi$, and $n_z$
can be obtained by the Hamilton equations ($\dot{p} = -\delta H/\delta q$
and $\dot{q} = \delta H/\delta p$)
\begin{mathletters}
\label{jj}
\begin{equation}
\frac{d n_z(q)}{dt} = -\frac{\delta H_f}{\delta \frac{\hbar}{2}\varphi(-q)} =
-\frac{2}{\hbar} K_{\varphi\varphi}(q) \varphi(q),
\label{jja}
\end{equation}  
\begin{equation}
\frac{d \varphi(q)}{dt} = \frac{\delta H_f}{\delta \frac{\hbar}{2}n_z(-q)} =
\frac{2}{\hbar} K_{zz}(q) n_z(q),
\label{jjb}
\end{equation}  
\end{mathletters}
where $n_z$ is the momentum density associated with the field variable 
$\varphi$.
In the spin coherent representation, $n_z$ and $n_+$ are given by
$-i\hbar\partial / \partial\varphi$, and $e^{i\varphi}$ respectively.
Then it is easy to check the transformation
$n_z \rightarrow \partial_t \varphi$ can be
obtained by canonical quantization, i.e., $n_z$ can be considered as
the momentum density conjugate to field variable $\varphi$. 
Making the derivative with respect to time we find the equation of motion: 
$\ddot{n}_z = -(2/\hbar) K_{\varphi\varphi}(q) \dot{\varphi}(q) =
-(2/\hbar)^2 K_{\varphi\varphi}(q)K_{zz}(q) n_z$, where $\dot{\varphi}
\equiv d\varphi/dt$. This clearly gives 
\begin{equation}
\hbar\omega = 2 \sqrt{K_{\varphi\varphi}(q)K_{zz}(q)}.
\label{spinwaves}
\end{equation}
Eqs.(\ref{jja}-\ref{jjb}) are similar to the coupled 
Josephson junction relations in superconductivity.

\section{Acknowledgement}
We are grateful to Homayoun Hamidian who was involved in the early
stage of this study.
Useful conversation with Allan MacDonald, Miguel Martin-Delgado
and Kieran Mullen is acknowledged.
Work at the University of Oklahoma was supported by the NSF under grant
No. EPS-9720651 and a grant from the Oklahoma State Regents for Higher
Education.  

\end{multicols}
\end{document}